# Determination of bimolecular recombination constants in organic double-injection devices using impedance spectroscopy

Makoto Takada,[1] Takahiro Mayumi,[1] Takashi Nagase,[1,2] Takashi Kobayashi,[1,2] and Hiroyoshi Naito[1,2, a]

[1] *Department of Physics and Electronics, Osaka Prefecture University, Sakai, 599-8531, Japan*

[2] *The Research Institute for Molecular Electronic Devices, Osaka Prefecture University, 1-1 Gakuen-cho, Naka-ku, Sakai, 599-8531, Japan*

A method for determination of the bimolecular recombination constant in working double-injection diodes such as organic light-emitting diodes (OLEDs) and organic photovoltaics (OPVs) using impedance spectroscopy is proposed. The proposed method is based on the theory that was developed to interpret the negative capacitance that has previously been observed in both OLEDs and OPVs. The determination of the bimolecular recombination constants is demonstrated using working polymer light-emitting diodes based on poly(9,9-dioctylfluorene-alt-benzothiadiazole). These impedance spectroscopy measurements thus allow us to determine the required bimolecular recombination coefficients along with the electron and hole drift mobilities in working organic double-injection diodes.

Organic double-injection devices, including organic light-emitting diodes (OLEDs) and organic photovoltaics (OPVs), have attracted considerable attention as optoelectronic applications of organic materials that offer several unique features, including flexibility, light weight and printability. It is essential to characterize the carrier transport properties (e.g., carrier drift mobility, localized state distributions, deep trapping lifetime, and bimolecular recombination constant) of organic semiconductors to understand the device physics of OLEDs and OPVs and aid in device structure design. The bimolecular recombination constant, which governs the recombination rate and the recombination zone, is one of the most important physical quantities that characterizes the transport properties and is closely related to both the external quantum efficiency of OLEDs and the power conversion efficiency of OPVs.

---

[a] Author to whom correspondence should be addressed. Electronic mail: naito@pe.osakafu-u.ac.jp

The bimolecular recombination process in low-mobility semiconductors (organic semiconductors are generally low-mobility semiconductors) is known to be Langevin recombination [1]. The Langevin recombination constant $\gamma_L$ is given by

$$\gamma_L = \frac{q}{\varepsilon}(\mu_n + \mu_p), \qquad (1)$$

where $q$ is the elementary charge, $\varepsilon$ is the dielectric constant, and $\mu_n$ and $\mu_p$ are the electron mobility and the hole mobility, respectively. The validity of Langevin's theory was studied in anthracene by Karl and Sommer in 1971 [2]. However, recent reports found that the bimolecular recombination constants of organic semiconducting materials used in OLEDs and OPVs are much lower than the Langevin recombination constant [3].

The bimolecular recombination constant of an organic semiconductor is estimated by fitting its current-voltage characteristics based on a double-injection steady-state space-charge-limited current (SCLC) theory to experimental results obtained in OLEDs [4]. However, experimental current-voltage characteristics cannot always be fitted using steady-state SCLC theory and prior knowledge of electron and hole mobilities is required to analyze the steady-state SCLC theory. To date, several transient electro-optical techniques have been proposed to determine the recombination constant, including time-of-flight (TOF)-based methods [5, 6], the transient photovoltage method [7], and the photo-induced current and charge extraction by linearly increasing voltage (photo-CELIV) method [8, 9]. However, TOF techniques require semiconducting layers with a minimum thickness of 1 μm to ensure a well-defined flight distance. In addition, the transient photovoltage method is only applicable to the characterization of OPVs, while the photo-CELIV method cannot be used to determine the bimolecular recombination constant in organic semiconductors that exhibit dispersive transport.

Impedance spectroscopy (IS) is a powerful tool for study of transport processes in organic semiconductors. The major advantages of IS measurement are that (i) IS measurements are fully automatic; (ii) IS measurements can be performed in thin-film devices with thicknesses of ~100 nm, which is comparable to the active layer thicknesses of both OLEDs and OPVs; and (iii) they allow transport properties to be studied in working devices. Characterization of the transport properties of thin organic semiconducting layers with thicknesses of ~100 nm is essential to studies of the device physics of OLEDs and OPVs because the optical and electronic properties of these semiconductors are thickness-dependent [10]. We have previously demonstrated simultaneous determination of the drift mobilities, localized state distributions, and deep trapping lifetimes in single-injection devices, which are also known as electron-only or hole-only devices (EODs or HODs) [11–13]. We have also determined the electron and hole drift mobilities simultaneously in double-injection devices [14].

In this letter, we propose a method using IS to determine the bimolecular recombination constants in working double-injection devices. Use of IS allows us to determine the electron and hole mobilities and the bimolecular recombination constants in working double-injection devices simultaneously. We validate the proposed method using device simulations and demonstrate the applicability of this method to determination of the bimolecular recombination constants of working polymer LEDs (PLEDs).

In double-injection devices such as OLEDs and OPVs, electrons and holes are injected into the light-emitting layers and the bulk heterojunction layers, respectively. In previous work [15], we derived an analytical expression for the small-signal impedance of double-injection devices in the presence of bimolecular recombination to explain the negative capacitance that was observed in the capacitance-frequency characteristics of both OLEDs and OPVs [16, 17]. In addition to our previous work, the origins of negative capacitance have been extensively discussed in literature (carrier injection through interfacial states [18], trap-assisted monomolecular recombination [19], self heating [20, 21], and energetic disorder [22]).

In the following, we present an analytical expression for double-injection devices. We have assumed a double-injection SCLC condition (i.e., where electrons and holes are injected into an organic semiconductor thin film). The device structure studied here is composed of an organic semiconductor thin film with planar geometry sandwiched between two electrodes. We also assumed that there were no localized states within the band gap. The basic equations that govern double injection in organic semiconductors are the time ($t$)-dependent current density equation, the continuity equations and Poisson's equation:

$$J = q\mu_n(n_T + \delta_n)F + q\mu_p(p_T + \delta_p)F + \varepsilon\frac{\partial F}{\partial t} = q\mu_p(b+1)nF + \varepsilon\frac{\partial F}{\partial t}, \qquad (2)$$

$$\frac{\partial n}{\partial t} = \frac{1}{q}\frac{\partial J_n}{\partial x} - \beta np, \quad \frac{\partial p}{\partial t} = -\frac{1}{q}\frac{\partial J_p}{\partial x} - \beta np, \qquad (3)$$

$$\frac{\partial F}{\partial x} = \frac{q}{\varepsilon}(\delta_p - \delta_n), \qquad (4)$$

where $J$ is the total current density, $n_T$ and $p_T$ are the electron and hole densities at thermal equilibrium, respectively, and $\delta_n$ and $\delta_p$ are the injected electron and hole densities, respectively. $F$ is the electric field, $b$ is the ratio of electron mobility to hole mobility (i.e., $b=\mu_n/\mu_p$), $n$ and $p$ are the total electron and hole densities (where $n=n_T+\delta_n$ and $p=p_T+\delta_p$), respectively, $x$ is the position along the current flow direction, and $\beta$ is the bimolecular recombination constant. The mobility is assumed to be both field- and time-independent under small ac perturbation. Diffusion of the charge carriers is also neglected. These assumptions are reasonable when the applied voltage is higher than several $kT/q$. We further assume that $n_T+\delta_n \approx \delta_n$ and $p_T+\delta_p \approx \delta_p$. This assumption is valid because organic semiconductors are generally insulating-type semiconductors, and $n_T$ and $p_T$ are thus extremely low when compared with $\delta_n$ and $\delta_p$, respectively.

From Eqs. (2) and (3), we have

$$\frac{b+1}{\mu_n}\frac{\partial n}{\partial t} = -\frac{b+1}{\mu_n}\beta np + (n_T - p_T)\frac{\partial F}{\partial x}. \tag{5}$$

We consider small ac signal perturbation for the impedance analysis, as follows:

$$J = J_0 + J_1\exp(j\omega t),\ n = n_0 + n_1\exp(j\omega t),\ F = F_0 + F_1\exp(j\omega t), \tag{6}$$

where $j$ is the imaginary unit and $\omega\ (=2\pi f)$ is the angular frequency. We then obtain the ac small-signal impedance $Z$ from Eq. (2)-(6) as

$$Z = 3R_0 \sum_{k=0}^{\infty}\frac{1}{k+3}\frac{\left(-j\omega\frac{3}{2}R_0 C_0\right)^k\left(2+\frac{j\omega}{\beta n_0}\right)^{k+1}}{\left(3+\frac{j\omega}{\beta n_0}\right)\left(3+\frac{j\omega}{\beta n_0}+1\right)\cdots\left(3+\frac{j\omega}{\beta n_0}+k\right)}, \tag{7}$$

where $R_0$ is the resistance of the semiconductor thin film (i.e., $R_0=V_0/I_0$), $V_0$ is the applied dc voltage, $I_0$ is the dc current (where $I_0=J_0 S$), $S$ is the active area, $C_0$ is the geometrical capacitance (where $C_0 = \varepsilon S/d$). In previous work [15], we showed from analysis of Eq. (7) that the negative capacitance occurs when the bimolecular recombination constant is lower than the Langevin recombination constant and when the value of the mobility balance $b$ is $>10^{-4}$.

The method to determine the bimolecular recombination constants proposed here is based on an analytical solution for the small-signal ac impedance of double-injection space-charge-limited diodes [Eq. (7)]. At sufficiently low frequencies (where $\omega\Theta<<1$), the first term in Eq. (7) is dominant, and thus the diode impedance becomes

$$Z = R_0\frac{2+j\omega/\beta n_0}{3+j\omega/\beta n_0}. \tag{8}$$

The imaginary part of Eq. (8) is given by

$$\text{Im}[Z] = R_0\frac{\omega/\beta n_0}{9+\omega^2/\beta^2 n_0^2}. \tag{9}$$

By differentiating Im[Z] with respect to $\omega$, we obtain $\omega_0=3\beta n_0$ at $d\text{Im}[Z]/d\omega = 0$, and thus obtain a $\beta$ value from $f_0$ at the maximum of Im[Z]:

$$\beta = \frac{\omega_0}{3n_0} = \frac{2\pi}{3n_0}f_0. \tag{10}$$

$f_0$ is observed when $\beta$ is smaller than $\gamma_L$, i.e., when the capacitance is negative [15]. When the capacitance is not negative, $\beta$ cannot be determined and is greater than or equal to $\gamma_L$. Here, we estimate the value of $n_0$ from the current density-voltage (*J-V*) characteristics of the working double-injection devices using

$$J = q(\mu_n + \mu_p)n_0 F. \tag{11}$$

We examined the validity of estimation of $n_0$ based on Eq. (11) using a device simulator (Atlas, Silvaco). In the simulation, as per the analytical derivation, we assume a structure composed of a thin film sandwiched between two planar electrodes that contains no localized states within the band gap. The $n_0$ value estimated from the simulated *J-V* characteristics using Eq. (11) is almost the same as that obtained from the spatial average of the carrier-density profile in the organic semiconductor, even under space-charge-limited conduction conditions (see Figs. S1 and S2, supplementary material).

Equation (11) was derived under the assumption of ohmic contact for both electrons and holes (meaning that $n \approx p$). However, in general, the fabrication of perfectly ohmic contacts on an organic semiconductor is extremely difficult [23]. In addition, Eq. (7), which was derived theoretically above, is not a function of *b*. To demonstrate the validity of the method proposed above for double-injection devices, the effects of both the Schottky injection barrier height and *b* on the determination of $\beta$ from the IS measurements were studied numerically using a device simulator. The physical quantities that were used in the simulations were appropriate for organic semiconductors [14, 24–26].

The Schottky injection barrier height for holes, denoted by $\phi_p$, is varied from 0 to 0.2 eV. At a high injection barrier height ($\phi_p > 0.2$ eV), the inductive response (i.e., the negative capacitance) is not observed in the Im[*Z*]-*f* characteristics (see Fig. S3, supplementary material), thus indicating that the device can be regarded as a single-injection device. When $\phi_p < 0.2$ eV, the inductive response, and thus the peak in the Im[*Z*]-*f* characteristics, can be observed and the $\beta$ value can be extracted from the peak frequency, $f_0$. The $\beta$ value determined from the peak in the Im[*Z*]-*f* characteristics is almost the same as the input $\beta$ for $\phi_p < 0.2$ eV (see Fig. S4, supplementary material). We also examined the effects of mobility balance *b* on the determination of $\beta$. Here, $\mu_p$ was varied from $10^{-5}$ to $10^{-1}$ cm$^2$V$^{-1}$s$^{-1}$ while $\mu_n$ remained fixed at $10^{-3}$ cm$^2$V$^{-1}$s$^{-1}$. The $\beta$ value that was determined from the peak in the Im[*Z*]-*f* characteristics was also almost the same as the input $\beta$ value in the $\mu_n/\mu_p$ range from $10^{-2}$ to $10^2$ (see Fig. S5, supplementary material). The device simulation thus shows that the proposed method can be successfully applied to determine the bimolecular recombination constants in working OLEDs and OPVs [27].

We performed IS on PLEDs based on a green light-emitting polymer, poly(9,9-dioctylfluorene-alt-benzothiadiazole) (F8BT), to demonstrate the applicability of the proposed method to the determination of $\beta$ in organic double-injection devices. The PLED was based on an AZO/PEI/F8BT/MoO$_3$/Al configuration, where AZO is Al-doped ZnO and PEI is

poly(ethyleneimine). A patterned AZO glass (Geomatec, Yokohama, Japan) layer that was used as a cathode was cleaned using acetone, 2-propanol and an ultraviolet (UV)-ozone method. Subsequently, a thin PEI layer, acting as an electron-injection layer, was spun onto the AZO glass surface from an ethanol solution (0.1 wt.%, 2000 rpm, 30 s). The substrate was then annealed in the ambient atmosphere (5 min, 150°C). A 130-nm-thick F8BT layer was spun onto the PEI layer from a chlorobenzene solution to act as an emissive layer (0.75 wt.%, 800 rpm, 60 s). After deposition of the F8BT emissive layer, the substrates were dried at 80°C for 30 min. 10-nm-thick $MoO_3$ and 50-nm-thick Al layers were then thermally evaporated successively onto the F8BT emissive layer in a vacuum chamber at a base pressure of $10^{-3}$ Pa. Finally, the resulting PLEDs were encapsulated using epoxy. The above processes were carried out in a nitrogen-filled glove box (dew point: −80°C). The active area of these PLEDs was 4.0 $mm^2$. Their current efficiency was 7.4 cd/A at 3.8 V (100 $mA/cm^2$, 7700 $cd/m^2$), which is almost comparable to that of F8BT PLEDs described in the literature [28–30] (see Fig. S6, supplementary material).

The PLEDs were held in a probe station (TTP-4, Desert Cryogenics). The $J$-$V$ characteristics and impedance spectra were measured using a Solartron Modulab XM over the range from $10^0$ to $10^6$ Hz at 160–300 K.

The frequency dependence of the negative differential susceptance ($-\Delta B = -\omega[C(\omega) - C_{geo}]$) at various temperatures at $V_{dc}$ = 3.0 V in the F8BT-based PLEDs is shown in the inset of Fig. 1. Two $-\Delta B$ peaks are observed over the 170–240 K range, which were caused by transits of the electrons and holes. The carrier drift mobility $\mu_d$ can be determined based on the carrier transit time $t_t$ that was obtained from the frequency $f_{max}$ at the maxima of $-\Delta B$ [31].

$$\mu = \frac{4}{3}\frac{d^2}{t_t V} \approx \frac{4}{3} \times \frac{f_{max}}{0.72}\frac{d^2}{V}. \tag{12}$$

The device simulation study has shown that electron and hole drift mobilities can be determined simultaneously in double injection devices [32]. Simultaneous determination of the electron and hole mobilities is possible under the following four conditions [32]: 1) the hole and electron mobilities in double injection devices can be determined simultaneously when the mobilities of the two carriers differ by a factor of at least 20; 2) the injection barrier heights for both the anode and the cathode in double-injection devices must be low enough (<0.1 eV) to allow simultaneous injection of both electrons and holes at comparable rates; 3) the recombination rate between the electrons and holes must be in a range from $10^{-1}$ to $10^{-3}$ smaller than $\gamma_L$, and no transit time effects should be observed in the case of strong recombination (>$\gamma_L$); and 4) the trap concentrations must be sufficiently low to ensure that their contributions to the impedance spectra do not mask the susceptance peak that corresponds to the carrier with the lowest mobility. In Ref.14, 33, the simultaneous measurements of the electron and hole mobilities using Eq. (12) have been demonstrated in tris(8-hydroxyquinolinato) aluminum ($Alq_3$) double-injection diodes and poly(p-phenylene vinylene) (PPV) polymer LEDs.

Plots of $\mu_n$ and $\mu_p$ versus temperature under various electric fields in F8BT-based PLEDs are shown in Fig. 1. The $\mu_n$ and $\mu_p$ values of F8BT at 300 K and 130 KV cm$^{-1}$ are $1.5\times10^{-4}$ and $4.6\times10^{-5}$ cm$^2$V$^{-1}$s$^{-1}$, respectively, and are consistent with previously reported values for F8BT in the literature [34–36]. The $\mu_n$ and $\mu_p$ values are also consistent with those estimated from the steady-state current - voltage characteristics of F8BT EODs (AZO/PEI/F8BT/Ca/Al) and HODs (ITO/poly(3,4-ethylenedioxythiophene):poly(styrenesulfonate) (PEDOT:PSS)/F8BT/MoO$_3$/Al) using a theory of single-carrier SCLC with field-dependent mobility [37, 38].

The Im[$Z$]-$f$ characteristics in the PLEDs under various applied voltages are shown in Fig. 2. The electric field dependence of the value of $\beta$ in Fig. 3 is determined from the $f_0$ that was observed in Fig. 2 using the values of $n_0$ from Eq. (11), while the values of $\gamma_L$ that were determined from $\mu_n$ and $\mu_p$ in Fig. 1 using Eq. (1) are also shown in Fig. 3. The value of $\beta$ at 300 K and 130 KV cm$^{-1}$ is $9.8\times10^{-13}$ cm$^3$s$^{-1}$, which is approximately $10^2$ times smaller than $\gamma_L$. A value of $\beta$ similar to that reported in this study has previously been reported in F8BT (where a simulation-based study of the performance of F8BT-based organic light-emitting field-effect transistors (OLETs) showed $\beta = 10^{-2}\gamma_L - 10^{-3}\gamma_L$ [39]). The mechanisms of these small recombination constants have been discussed and were attributed to the spatial separation of the electrons and holes [3, 40]. The random spatial fluctuations in the potential landscape in disordered materials can lead to spatially separated percolation networks for the electrons and holes, which consequently leads to a low average carrier recombination rate [3]. These smaller recombination constants for F8BT ($\beta \approx 10^{-2}\gamma_L$) can be explained by the amplitude of the potential fluctuations being 80 meV.

High current efficiency has been reported in F8BT-based PLEDs with thick F8BT emission layers (~1 μm) [41]. The maximum current efficiencies of the F8BT PLEDs were 12, 19, 21, and 23 cd/A for emission layer thicknesses of 200, 350, 870, and 1200 nm, respectively, where the PLED device configuration was ITO/ZnO/Cs$_2$CO$_3$/F8BT/MoO$_3$/Au [41]. We speculate that this emission layer thickness dependence of the current efficiency is due to the weak bimolecular recombination in F8BT that was observed above, and we performed device simulations to examine the emission layer thickness dependence of the current efficiency. The value of $\beta$ was varied from $10^{-3}\gamma_L$ to $10^0\gamma_L$ cm$^3$s$^{-1}$, and the emission layer thickness was varied from 50 to 500 nm. A greater thickness dependence for the current efficiency was observed at smaller $\beta$ values (see Fig. S7, supplementary material). The device simulation results are consistent with the experimental results [41] because weaker bimolecular recombination and unbalanced electron and hole mobilities, as occur in the F8BT case, lead to wider recombination zones in the PLEDs. The F8BT thickness dependence of the current efficiency [41] thus serves as another indication that $\beta$ is much lower than $\gamma_L$.

In conclusion, we propose a method to determine the bimolecular recombination constants in organic double-injection devices using IS, which also allows us to determine the electron and hole drift mobilities in these double-injection devices. We demonstrate the validity of the proposed method through device simulations. We demonstrate the applicability of the proposed method to determination of the bimolecular recombination constant $\beta$ in PLEDs based on F8BT. The value of $\beta$ for F8BT at 300 K and 130 KV cm$^{-1}$ is $9.8\times10^{-13}$ cm$^3$s$^{-1}$, which is $10^2$ times smaller than the Langevin recombination constant $\gamma_L$. The F8BT layer thickness dependence of the current efficiency in F8BT-based PLEDs, as reported in the literature, is related to these low $\beta$ values. The method proposed here is applicable to simultaneous determination of the electron and hole mobilities and the bimolecular recombination constants in organic double-injection devices such as PLEDs and OPVs. These physical constants are very useful for both device design and understanding of the related device physics.

Please see the supplementary material for details of the validity of the estimation of the free carrier density, the effects of the carrier injection barriers and carrier mobility balances on determination of the bimolecular recombination constants based on device simulations, the device characteristics of the F8BT PLEDs, and the effects of the emission layer thickness on the current efficiency for various values of the bimolecular recombination constants based on device simulations.

This work was partly supported by JSPS KAKENHI under Grant Number JP17H01265, by the Murata Science Foundation under Grant Number H29RS72, and by the ICOM Foundation, Japan. The authors would like to thank Sumitomo Chemical Company, Ltd. for supplying the F8BT, and Nippon Shokubai Company, Ltd. for supplying the PEI. We thank David MacDonald, MSc, from Edanz Group (www.edanzediting.com/ac) for editing a draft of this manuscript.

(Y. Liu, Y. Gao, B. Xu, P. H. M. van Loosdrecht, and W. Tian, Org. Electron. **38**, 8 (2016)), transient absorption spectroscopy (C. G. Shuttle, B. O'Regan, A. M. Ballantyne, J. Nelson, D. D. C. Bradley, and J. R. Durrant, Phys. Rev. B **78**, 113201 (2008)), time-of-flight (A. Pivrikas, G. Juška, A. J. Mozer, M. Scharber, K. Arlauskas, N. S. Sariciftci, H. Stubb, and R. Österbacka, Phys. Rev. Lett. **94**, 176806 (2005)), and *J-V* characteristics (G.-J. A. H. Wetzelaer, N. J. Van der Kaap, L. J. A. Koster, P. W. M. Blom, Advanced Energy Materials, **3**, 1130 (2013)) in P3HT:PCBM OPVs.

**FIGURES**

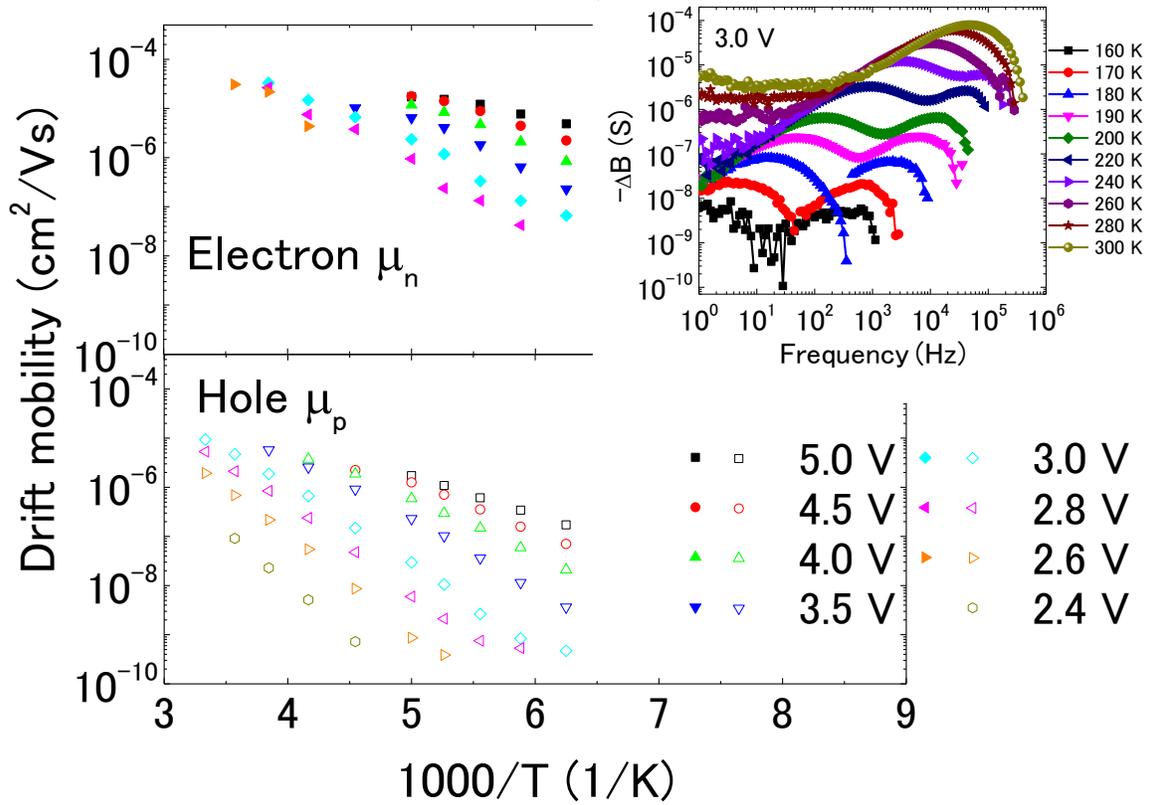

Fig. 1. Plots of electron and hole drift mobilities ($\mu_n$ and $\mu_p$, respectively) versus temperature with various electric fields in F8BT-based PLEDs. At 300 K, $\mu_n$ and $\mu_p$ are estimated to be $10^{-4}$ and $10^{-5}$ cm$^2$V$^{-1}$s$^{-1}$, respectively. The inset shows the frequency dependence of the negative differential susceptance ($-\Delta B = -\omega[C(\omega) - C_{\text{geo}}]$) for different temperatures at $V_{\text{dc}}$ = 3.0 V in F8BT-based PLEDs. Two $-\Delta B$ peaks are observed in the 170−240 K range; they were caused by electron and hole transit-time effects.

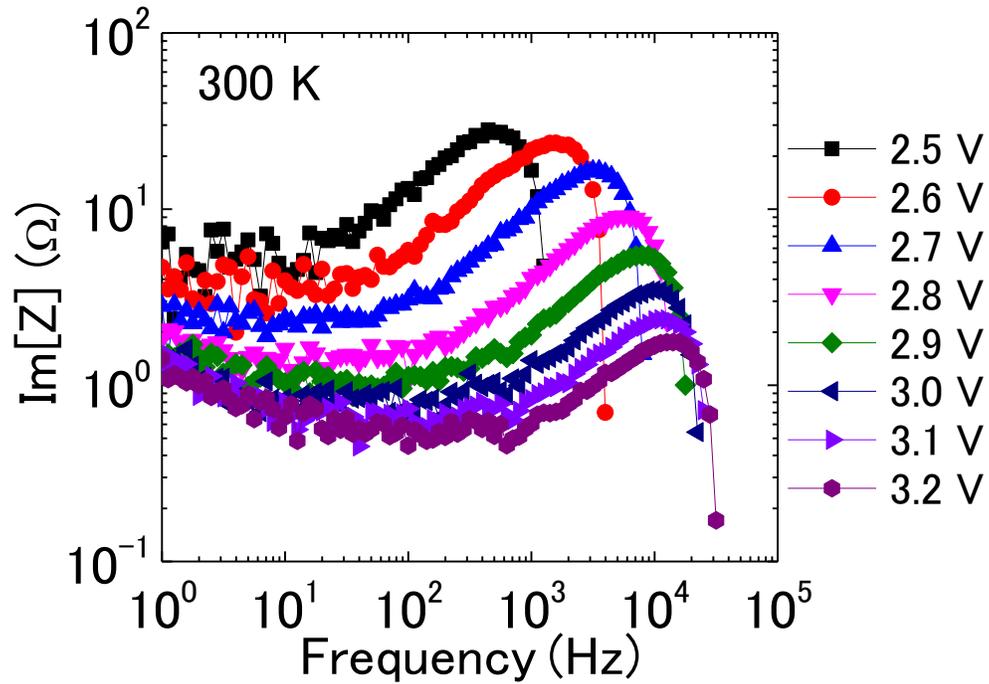

Fig. 2. Frequency dependence of imaginary component of impedance under various applied voltages in F8BT-based PLEDs.

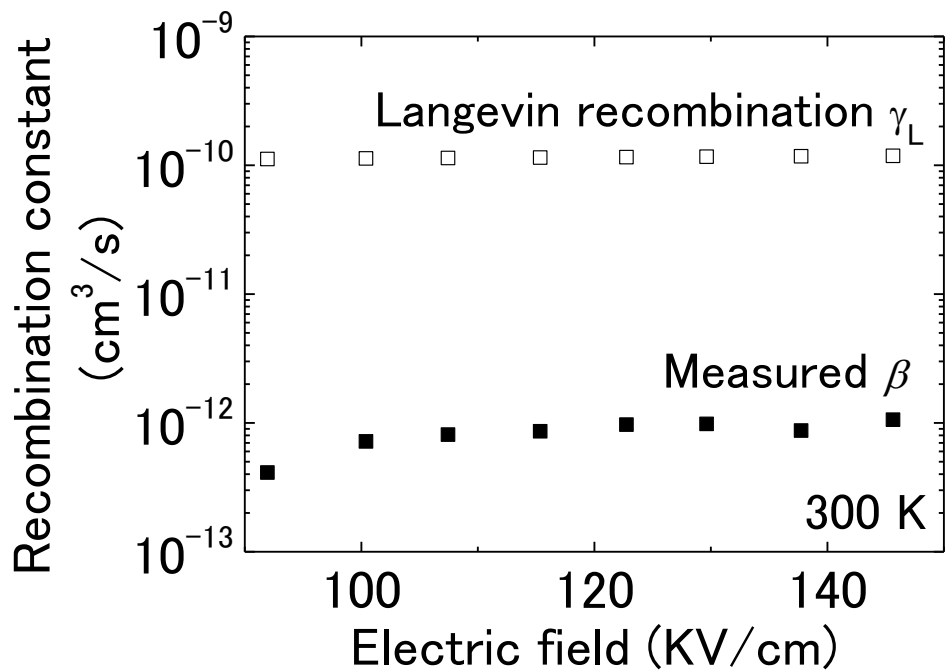

Fig. 3. Electric field dependence of recombination constant of F8BT determined from the frequency of the maximum of Im[$Z$] in Fig. 2 (closed symbols) and the Langevin recombination constant (open symbols).

# Supplementary Material

# Determination of bimolecular recombination constants in organic double-injection devices using impedance spectroscopy

Makoto Takada,[1] Takahiro Mayumi,[1] Takashi Nagase,[1,2] Takashi Kobayashi,[1,2] and Hiroyoshi Naito[1,2,2)]

[1] *Department of Physics and Electronics, Osaka Prefecture University, Sakai, 599-8531, Japan*

[2] *The Research Institute for Molecular Electronic Devices, Osaka Prefecture University, 1-1 Gakuen-cho, Naka-ku, Sakai, 599-8531, Japan*

---

[2] Author to whom correspondence should be addressed. Electronic mail: naito@pe.osakafu-u.ac.jp

## S1. Estimation of free carrier density in working devices

We demonstrated the validity of the proposed method for estimation of the free carrier density in double-injection devices. The following physical quantities were used in the calculations: $N_C = N_V = 2.5\times10^{19}$ cm$^{-3}$, $\varepsilon_r = 3$, $E_{CB} = 3.1$ eV, $E_{VB} = 5.8$ eV, $d = 100$ nm, $V = 0\text{–}10$ V, $T = 300$ K, $\phi_n = \phi_p = 0$ eV, $\mu_n = \mu_p = 10^{-3}$ cm$^2$V$^{-1}$s$^{-1}$, and $\beta = \gamma_L$; here, $N_{C(V)}$, $\varepsilon_r$, $E_{CB(VB)}$, $\phi_{n(p)}$, $\beta$, and $\gamma_L$ are the effective density of states of the conduction band (valence band), the relative permittivity, the energy at the bottom of the conduction band (top of the valence band), the Schottky injection barrier height for electrons (holes), the bimolecular recombination constant, and the Langevin recombination constant, respectively. The carrier density values that were obtained from the carrier density profile and the corresponding value estimated from the current density-voltage ($J$-$V$) characteristics plotted versus the electric field are shown in Fig. S1. The carrier density values that were estimated from the $J$-$V$ characteristics were almost the same as the value that was obtained from the carrier density profile.

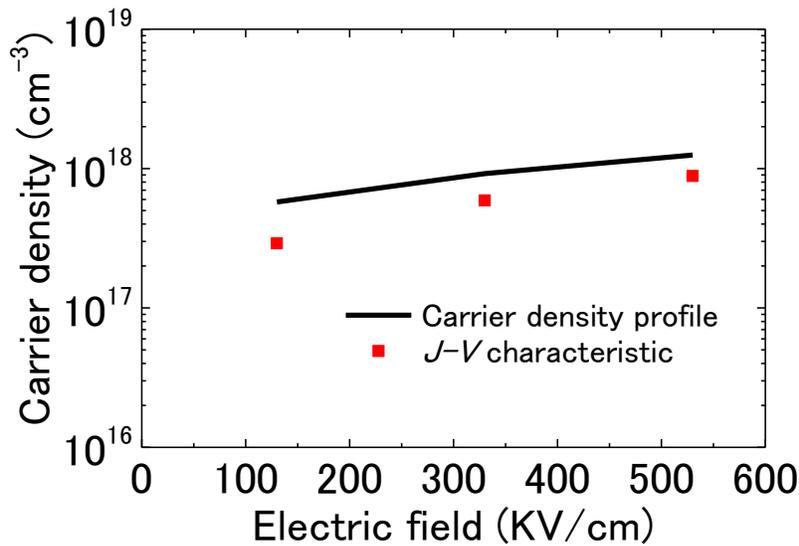

Fig. S1. Carrier density values obtained from the carrier density profile (solid line) and from the $J$-$V$ characteristics using Eq. (15) (dots) versus the electric field, as calculated using a device simulator. The following physical quantities were used in the calculations: $N_C = N_V = 2.5\times10^{19}$ cm$^{-3}$, $\varepsilon_r = 3$, $E_{CB} = 3.1$ eV, $E_{VB} = 5.8$ eV, $d = 100$ nm, $V_{dc} = 4, 6,$ and $8$ V, $T = 300$ K, $\phi_n = \phi_p = 0$ eV, $\mu_n = \mu_p = 10^{-3}$ cm$^2$V$^{-1}$s$^{-1}$, and $\beta = \gamma_L$.

We also examined the effects of $\beta = \alpha\gamma_L$ on estimation of the carrier density. The values of $\alpha$ were varied from $10^{-5}$ to $10^0$. The plots of the carrier density obtained from the carrier density profile and the carrier density estimated from the $J$-$V$ characteristics versus $\beta$ are shown in Fig. S2. The carrier density values that were estimated from the current density-voltage ($J$-$V$) characteristics were almost the same as that obtained from the carrier density profile at the different values of $\beta$.

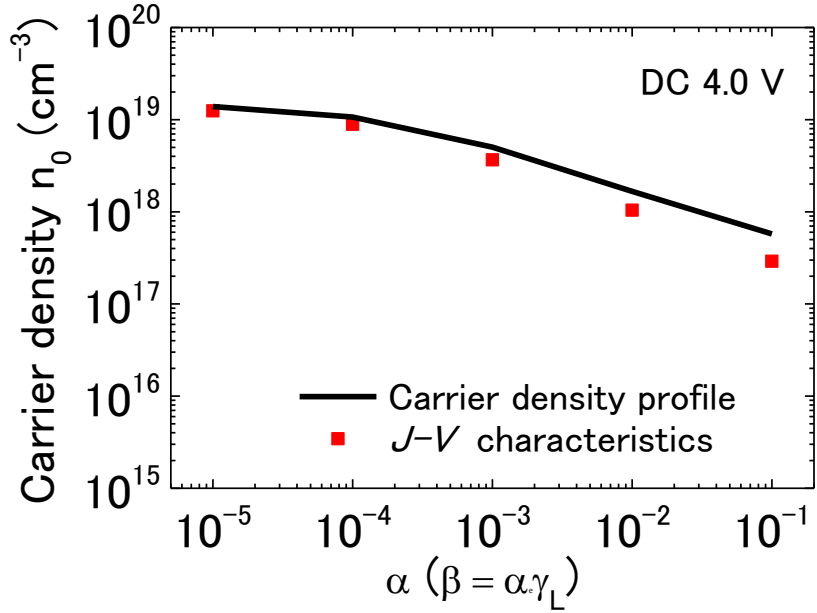

Fig. S2. Plots of carrier density obtained from the carrier density profile (solid line) and from the *J-V* characteristics using Eq. (15) (dots) versus $\alpha$ calculated using a device simulator. The following physical quantities were used in the calculations: $N_C = N_V = 2.5 \times 10^{19}$ cm$^{-3}$, $\varepsilon_r = 3$, $E_{CB} = 3.1$ eV, $E_{VB} = 5.8$ eV, $d = 100$ nm, $V_{dc} = 4.0$ V, $T = 300$ K, $\phi_n = \phi_p = 0$ eV, $\mu_n = \mu_p = 10^{-3}$ cm$^2$V$^{-1}$s$^{-1}$, and $\beta = 10^{-5}\gamma_L - 10^0\gamma_L$ ($\alpha = 10^{-5} - 10^0$).

## S2. Validity of the method for determination of bimolecular recombination constants

First, we examine the effects of the injection barrier on determination of the bimolecular recombination constants in double-injection devices. $\phi_p$ is varied from 0 to 0.2 eV and we assume injection by the Schottky injection mechanism. The frequency dependences of the imaginary component of the impedance (Im[*Z*]) at $V_{dc} = 4.0$ V with different hole-injection barriers calculated using a device simulator are shown in Fig. S3. At the high injection barrier height ($\phi_p > 0.2$ eV), the inductive response (i.e., the negative capacitance) is not observed in the Im[*Z*]-*f* characteristics, indicating that the device can be regarded as a single-injection device.

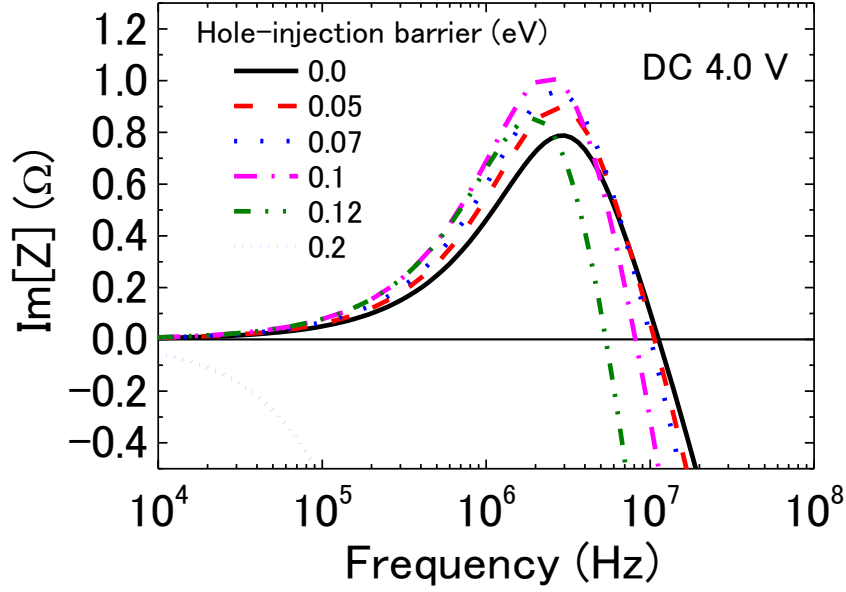

Fig. S3. Frequency dependences of the imaginary component of impedance (Im[$Z$]) at $V_{dc}$ = 4.0 V with different hole-injection barriers as calculated using a device simulator. The following physical quantities were used in the calculations: $N_C$ = $N_V$ = 2.5×10$^{19}$ cm$^{-3}$, $\varepsilon_r$ = 3, $E_{CB}$ = 3.1 eV, $E_{VB}$ = 5.8 eV, $d$ = 100 nm, $V_{dc}$ = 4.0 V, $T$ = 300 K, $\phi_n$ = 0, $\phi_p$ = 0 – 0.2 eV, $\mu_n$ = $\mu_p$ = 10$^{-3}$ cm$^2$V$^{-1}$s$^{-1}$, and $\beta$ = 10$^{-11}$ cm$^3$s$^{-1}$. At the high injection barrier height ($\phi_p$ > 0.2 eV), the inductive response is not observed in the Im[$Z$]-$f$ characteristics.

At $\phi_p$ < 0.2 eV, the inductive response, and thus a peak in the Im[$Z$]-$f$ characteristics, is observed and the value of $\beta$ can then be extracted from the peak frequency, $f_0$. The extracted value of $\beta$ is almost the same as the input $\beta$ (Fig. S4). Almost the same results as those shown in Fig. S4 were obtained in the case where carrier injection barriers were present at both the anode and the cathode ($\phi_p$ [= $\phi_n$] < 0.2 eV).

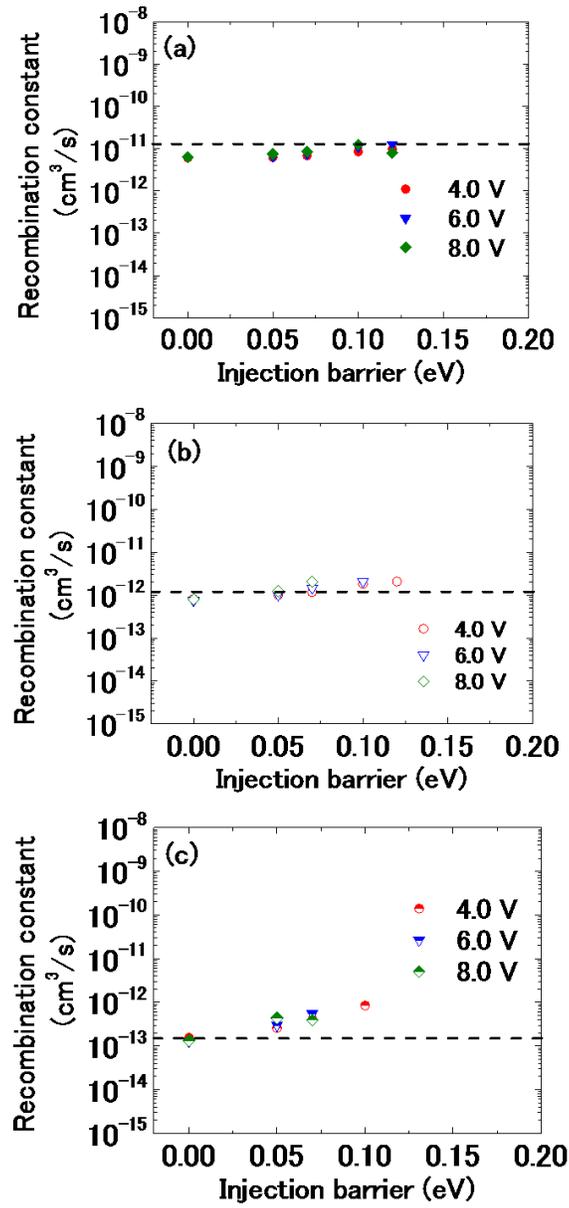

Fig. S4. Plots of bimolecular recombination constant versus hole injection barrier height for different input bimolecular recombination constants $\beta_{input}$: (a) $\beta_{input} = 10^{-11}$ cm$^3$s$^{-1}$, (b) $\beta_{input} = 10^{-12}$ cm$^3$s$^{-1}$, and (c) $\beta_{input} = 10^{-13}$ cm$^3$s$^{-1}$. The following physical quantities were used in the calculations: $N_C = N_V = 2.5 \times 10^{19}$ cm$^{-3}$, $\varepsilon_r = 3$, $E_{CB} = 3.1$ eV, $E_{VB} = 5.8$ eV, $d = 100$ nm, $V_{dc} = 4.0$ V, $T = 300$ K, $\phi_n = 0$, $\phi_p = 0 - 0.2$ eV, $\mu_n = \mu_p = 10^{-3}$ cm$^2$V$^{-1}$s$^{-1}$, and $\beta = 10^{-13} - 10^{-11}$ cm$^3$s$^{-1}$.

Second, we examine the effects of the mobility balance $b$ on the determination of $\beta$. $\mu_p$ is varied from $10^{-5}$ to $10^{-1}$ cm$^2$V$^{-1}$s$^{-1}$, $\mu_n$ is fixed at $10^{-3}$ cm$^2$V$^{-1}$s$^{-1}$, and the other physical quantities had the same values as those mentioned above. The extracted $\beta$ value is almost the same as the input $\beta$ value in the range of $\mu_n/\mu_p = 10^{-2}-10^2$ (Fig. S5).

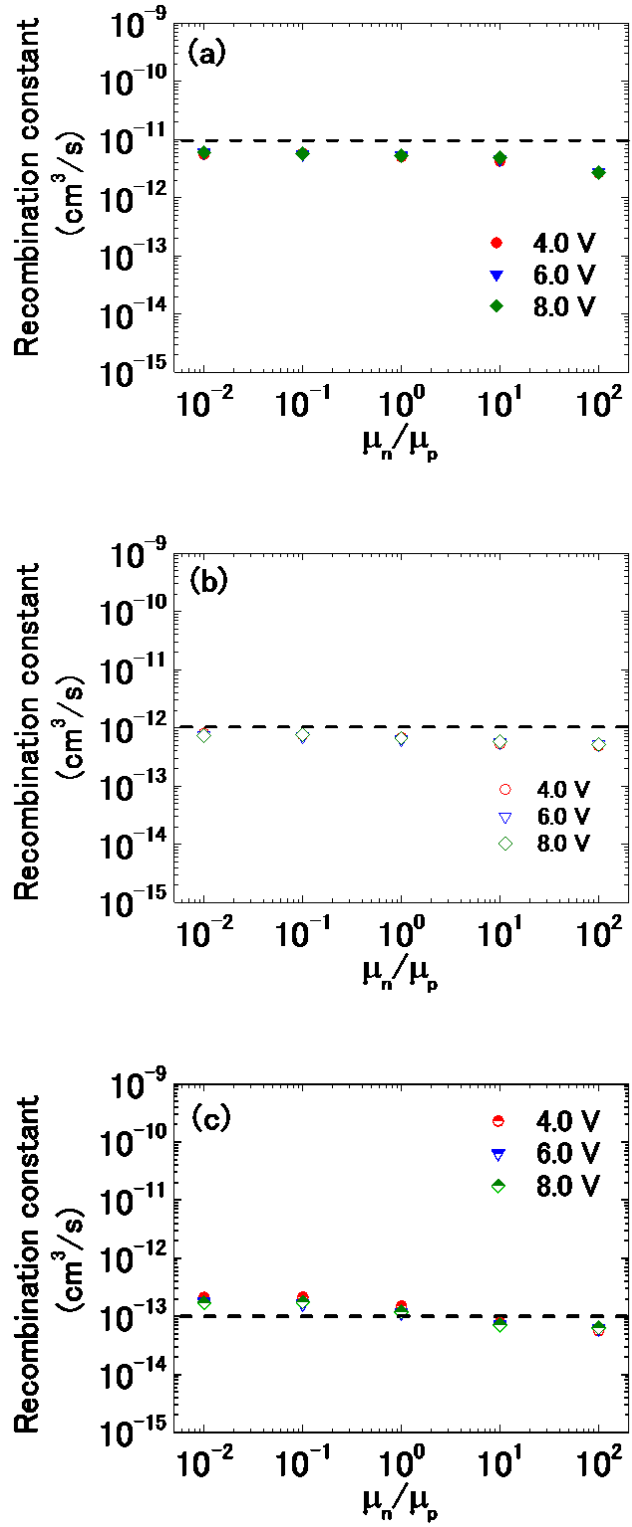

Fig. S5. Plots of bimolecular recombination constant versus electron and hole mobility balance for different values of input bimolecular recombination constant $\beta_{input}$: (a) $\beta_{input} = 10^{-11}$ cm$^3$s$^{-1}$, (b) $\beta_{input} = 10^{-12}$ cm$^3$s$^{-1}$, and (c) $\beta_{input} = 10^{-13}$ cm$^3$s$^{-1}$. The following physical quantities were used in the calculations: $N_C = N_V = 2.5 \times 10^{19}$ cm$^{-3}$, $\varepsilon_r = 3$, $E_{CB} = 3.1$ eV, $E_{VB} = 5.8$ eV, $d = 100$ nm, $V_{dc} = 4.0$ V, $T = 300$ K, $\phi_n = \phi_p = 0$ eV, $\mu_n = \mu_p = 10^{-3}$ cm$^2$V$^{-1}$s$^{-1}$, and $\beta = 10^{-13} - 10^{-11}$ cm$^3$s$^{-1}$.

## S3. Device performance of PLEDs

The *J-V* and luminescence-voltage characteristics of poly(9,9-dioctylfluorene-alt-benzothiadiazole) (F8BT)-based polymer light-emitting diodes (PLEDs) are shown in Fig. S6a. The PLEDs were fabricated to demonstrate the applicability of the method for determination of the bimolecular recombination constants, where the device has an Al-doped ZnO (AZO)/poly(ethyleneimine) (PEI)/F8BT/MoO$_3$/Al configuration. The luminescence at 4.5 V is 15,000 cd m$^{-2}$ (at 240 mA cm$^{-2}$). The current efficiency-current density characteristics of the PLEDs that were obtained from Fig. S6a are shown in Fig. S6b. The current efficiency at 100 mA cm$^{-2}$ is 7.4 cd A$^{-1}$.

The film thickness of PEI (poly(ethyleneimine)) is 1-5 nm, which is much thinner than that of F8BT (~100 nm). Thus, the dielectric relaxation due to the PEI layer is not observed in the impedance spectra of the F8BT OLEDs (10$^0$ – 10$^6$ Hz), and only the dielectric relaxation due to F8BT layer is observed in the impedance spectra. The roles of PEI for the improvement of OLED performance are discussed in Ref. S1, and are the electron-injection barrier lowering, the passivation of the surface states of AZO, and the blocking of excitons and holes at the AZO/F8BT interface [S1].

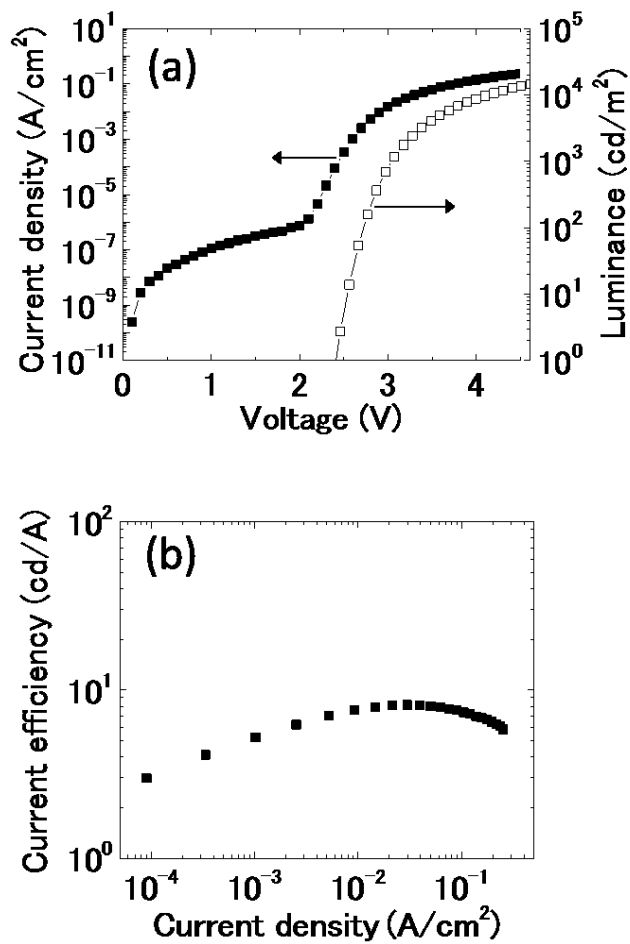

Fig. S6. (a) Current density-voltage and luminescence-voltage characteristics of poly(9,9-dioctylfluorene-alt-benzothiadiazole) (F8BT)-based polymer light-emitting diodes (PLEDs) with an Al-doped ZnO (AZO)/poly(ethyleneimine) (PEI)/F8BT/MoO$_3$/Al device configuration. (b) Current efficiency-current density characteristics of the PLEDs obtained from the results in Fig. S6a.

**S4. Influence of emission layer thickness on current efficiency at various bimolecular recombination constant values**

We examine the effects of the emission layer thickness on the current efficiency at different values of $\beta$. $\beta$ is changed from $10^{-5}\gamma_L$ to $10^0\gamma_L$ cm$^3$s$^{-1}$, and the emission layer thickness is varied from 50 to 500 nm. The emission-layer thickness dependences of the current efficiency at different values of $\beta$ are shown in Fig. S7; a greater thickness dependence for the current efficiency is found at smaller values of $\beta$.

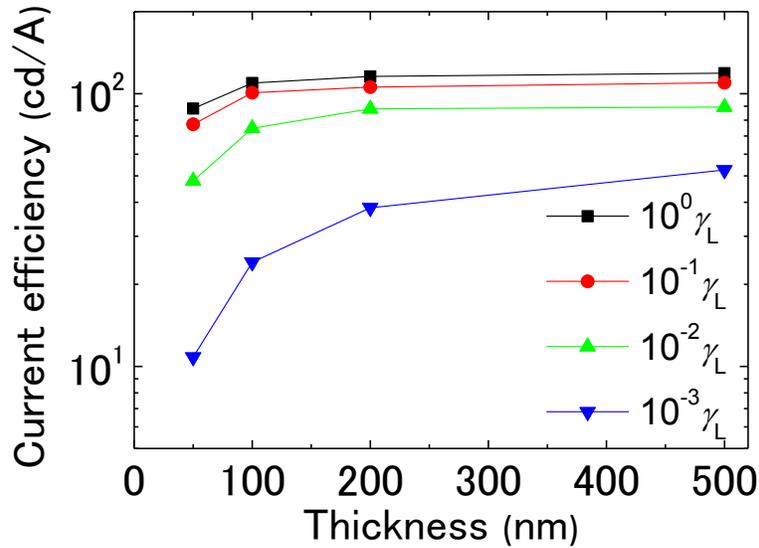

Fig. S7. Emission layer thickness dependences of current efficiency at various values of $\beta$ calculated using a device simulator. The following physical quantities were used in the calculations: $N_C = N_V = 2.5 \times 10^{19}$ cm$^{-3}$, $\varepsilon_r = 3$, $E_{CB} = 3.5$ eV, $E_{VB} = 5.9$ eV, $d = 50, 100, 200,$ and 500 nm, $T = 300$ K, $\phi_n = \phi_p = 0$ eV, $\mu_n = 10^{-4}$ cm$^2$V$^{-1}$s$^{-1}$, $\mu_p = 10^{-5}$ cm$^2$V$^{-1}$s$^{-1}$, and $\beta = 10^{-5}\gamma_L, 10^{-3}\gamma_L, 10^{-2}\gamma_L, 10^{-1}\gamma_L,$ and $10^0\gamma_L$.